\documentclass[notitlepage,a4paper,aps,prd,onecolumn,superscriptaddress,nofootinbib,longbibliography]{revtex4-1}
\usepackage{amsmath}
\usepackage{amsfonts}
\usepackage{amssymb}
\usepackage[latin1,utf8]{inputenc}
\usepackage[T1]{fontenc}
\usepackage[colorlinks=true]{hyperref}

\newcommand{\tp}[1]{\overset{\bullet}{#1}\vphantom{#1}}
\newcommand{\lc}[1]{\overset{\circ}{#1}\vphantom{#1}}

\newcommand{\DD}{\mathrm{D}}
\newcommand{\dd}{\mathrm{d}}

\begin{document}

\title{Scalar-torsion theories of gravity I:\\general formalism and conformal transformations}

\author{Manuel Hohmann}
\email{manuel.hohmann@ut.ee}
\affiliation{Laboratory of Theoretical Physics, Institute of Physics, University of Tartu, W. Ostwaldi 1, 50411 Tartu, Estonia}

\begin{abstract}
We discuss the most general class of teleparallel scalar-torsion theories of gravity in their covariant formulation. The only restrictions we impose are the invariance of the action under diffeomorphisms and local Lorentz transformations, as well as vanishing direct coupling of the matter fields to the teleparallel spin connection. In this general setting we discuss the implications of local Lorentz invariance and diffeomorphism invariance and derive the general structure of the field equations. Further, we show how different theories of this class are related to each other by conformal transformations of the tetrad and redefinitions of the scalar field. We finally show how the formalism can be generalized to an arbitrary number of scalar fields, and provide a few examples.
\end{abstract}

\maketitle

\section{Introduction}\label{sec:intro}
The most successful theory of gravity is general relativity (GR), as it has been proven to be consistent with a wide range of observations on different scales. It attributes the gravitational interaction to the curvature of the Levi-Civita connection of a (pseudo-)Riemannian geometry. Nevertheless, GR is challenged by a number of observations in cosmology, such as the accelerating phases in the early and late universe~\cite{Ade:2015xua,Ade:2015rim,Ade:2015lrj}. Further, it is challenged by its tension with quantum theory, as so far no fully conclusive theory of quantum gravity has been derived. These aspects have motivated the study of numerous alternative gravity theories. A large class of such theories is based on the idea that gravity is not attributed to the curvature of the torsion-free Levi-Civita connection, but to the torsion of a flat, i.e., curvature-free connection. This attempt is known as teleparallelism, since the flat connection allows for a path-independent parallel transport of tangent vectors~\cite{Moller:1961,Aldrovandi:2013wha,Maluf:2013gaa,Golovnev:2018red}. An interesting feature of teleparallel gravity is its possible interpretation as a gauge theory of the translation group, which brings it closer to other fundamental interactions which are modeled as gauge theories~\cite{Cho:1975dh,Cho:1976fr,Hayashi:1977jd}.

In its original formulation, teleparallel theories of gravity have been based on the Weitzenböck connection defined by a tetrad field such that the corresponding spin connection coefficients vanish. However, this formulation has been shown to be problematic due to the breaking of local Lorentz invariance~\cite{Li:2010cg,Sotiriou:2010mv}, as spurious degrees of freedom may appear~\cite{Li:2011rn,Ong:2013qja,Izumi:2013dca,Chen:2014qtl}. In order to solve these problems a covariant formulation has been adopted, in which a flat, but in general non-vanishing spin connection is introduced~\cite{Krssak:2015oua}. It has been shown that a naive variation of the action with respect to the spin connection is too restrictive, as it does not yield the desired field equations, and so either a different method of variation~\cite{Golovnev:2017dox} or different physical principles~\cite{Krssak:2017nlv} must be employed. Further, it has been shown for different classes of teleparallel theories that a constrained variation of the spin connection yields a set of field equations, which is identical to the antisymmetric part of the tetrad field equations~\cite{Golovnev:2017dox,Hohmann:2017duq}.

While the simplest and most well-studied teleparallel gravity theory is equivalent to general relativity, and hence known as the teleparallel equivalent of general relativity (TEGR), it provides an alternative starting point for modifications, in order to address the mentioned challenges to GR. A possible modification of TEGR is the introduction of a scalar field non-minimally coupled to torsion~\cite{Geng:2011aj}; this is similar to introducing a scalar field non-minimally coupled to curvature as a modification of GR. Numerous models of this type have been studied~\cite{Izumi:2013dca,Chakrabarti:2017moe,Otalora:2013tba,Jamil:2012vb,Chen:2014qsa}. However, it is common to employ the Weitzenböck connection in order to derive the torsion, which leads to the aforementioned potential issues arising from breaking local Lorentz invariance. These issues are addressed in a covariant formulation of scalar-torsion theories, and it has been shown that the relation between the field equations for the tetrad and the spin connection also holds in this case~\cite{Hohmann:2018rwf}.

The aim of this article is to generalize the aforementioned results to a larger class of teleparallel scalar-torsion theories of gravity, where we allow the scalar field to couple in an arbitrary way to all other (gravitational and matter) fields, and impose as only restrictions the invariance of the action under diffeomorphisms and local Lorentz transformations, as well as a vanishing coupling between the matter fields and the teleparallel spin connection. For this class of theories we derive the field equations and study the implications of the imposed restriction. In particular, we discuss the question whether the local Lorentz invariance relates the antisymmetric part of the tetrad field equations to the connection field equations, as has been proven for narrower classes of theories. We also discuss how different theories within this class can be related to each other by performing conformal transformations of the tetrad and scalar field redefinitions.

This article is the first in a series of three articles, and develops the general framework for the aforementioned class of scalar-torsion theories. Special subclasses will be discussed in two subsequent articles: a class we denote \(L(T, X, Y, \phi)\) and whose gravitational action depends on four scalar quantities derived from the gravitational variables is developed in the second article~\cite{Hohmann:2018dqh}; this class will be further restricted to a class which has properties similar to scalar-tensor gravity in the third and last article~\cite{Hohmann:2018ijr}.

The outline of this article is as follows. In section~\ref{sec:fields} we list the fundamental fields present in the class of theories we consider and introduce the notation we will use. These are then used in section~\ref{sec:action} to define the action, and set the notation for its variation. This variation is exploited in the following three sections: to derive the implications of local Lorentz invariance in section~\ref{sec:loclor} and diffeomorphism invariance in section~\ref{sec:diffeo}, and to derive the field equations in section~\ref{sec:feqs}. Conformal transformations are discussed in section~\ref{sec:conf}. We generalize our discussion to multiple scalar fields in section~\ref{sec:multi}. A few examples are given in section~\ref{sec:examples}. We end with a conclusion in section~\ref{sec:conclusion}.

\section{Fields in scalar-torsion gravity and their relations}\label{sec:fields}
The most important difference between teleparallel theories of gravity and theories based on Riemannian geometry is that the fundamental field defining the geometry is not a metric, but a coframe field \(\theta^a\), which can be expressed as a set of four one-forms
\begin{equation}
\theta^a = \theta^a{}_{\mu}\dd x^{\mu}
\end{equation}
labeled with a Lorentz index, which constitute a basis of the cotangent space \(T^*_xM\) for all spacetime points \(x \in M\). The corresponding dual bases of the tangent spaces \(T_xM\) constitute a frame field \(e_a\), which can be expressed as a set of four vector fields
\begin{equation}
e_a = e_a{}^{\mu}\partial_{\mu}\,.
\end{equation}
Since there is a one-to-one correspondence between frame and coframe fields, both a conventionally denoted by the term tetrad. Further, we consider a flat Lorentz spin connection \(\tp{\omega}^a{}_b\), which is likewise given by one-forms
\begin{equation}
\tp{\omega}^a{}_b = \tp{\omega}^a{}_{b\mu}\dd x^{\mu}\,.
\end{equation}
All quantities related to this connection, which is also called the teleparallel connection, will be denoted by a bullet (\(\bullet\)). Note that being Lorentzian implies antisymmetry, \(\tp{\omega}^{(ab)} = 0\), where indices are raised and lowered with the Minkowski metric \(\eta_{ab}\), while flatness implies vanishing curvature
\begin{equation}\label{eqn:zerocurv}
\tp{R}^a{}_b = \dd\tp{\omega}^a{}_b + \tp{\omega}^a{}_c \wedge \tp{\omega}^c{}_b = 0\,.
\end{equation}
In addition to the tetrad and the spin connection, we consider a scalar field \(\phi\). Finally, we consider a set of matter fields \(\chi^I\), which we label by an index \(I\). For simplicity of notation, we will assume that the matter fields \(\chi^I\) are also given by differential forms of rank \(k_I\). However, this is not essential for our derivation, and more general choices are possible.

For our calculations we will make use of the fact that, given a tetrad, spin connection \(\tp{\omega}^a{}_b\) is uniquely determined by its torsion
\begin{equation}\label{eqn:torsion}
T^a = \tp{\DD}\theta^a = \dd\theta^a + \tp{\omega}^a{}_b \wedge \theta^b\,,
\end{equation}
where \(\tp{\DD}\) denotes the covariant exterior derivative. In order to invert the relation~\eqref{eqn:torsion} and determine the spin connection from the torsion, it is most convenient to introduce the contortion
\begin{equation}\label{eqn:contortion}
K_{ab} = \frac{1}{2}\left(\iota_{e_b}\iota_{e_c}T_a + \iota_{e_c}\iota_{e_a}T_b - \iota_{e_a}\iota_{e_b}T_c\right)\theta^c\,,
\end{equation}
as well as the Levi-Civita connection
\begin{equation}\label{eqn:levicivita}
\lc{\omega}_{ab} = -\frac{1}{2}\left(\iota_{e_b}\iota_{e_c}\dd\theta_a + \iota_{e_c}\iota_{e_a}\dd\theta_b - \iota_{e_a}\iota_{e_b}\dd\theta_c\right)\theta^c\,,
\end{equation}
whose associated quantities we denote with an open circle (\(\circ\)), and which is uniquely defined by having vanishing torsion, \(\lc{\DD}\theta^a = 0\). It allows us to write the spin connection as
\begin{equation}\label{eqn:invtorsion}
\omega_{ab} = \lc{\omega}_{ab} + K_{ab} = \frac{1}{2}\left(\iota_{e_b}\iota_{e_c}T_a + \iota_{e_c}\iota_{e_a}T_b - \iota_{e_a}\iota_{e_b}T_c - \iota_{e_b}\iota_{e_c}\dd\theta_a - \iota_{e_c}\iota_{e_a}\dd\theta_b + \iota_{e_a}\iota_{e_b}\dd\theta_c\right)\theta^c\,.
\end{equation}
These are the geometric objects we will need in order to define the action for the class of theories we consider here. This will be done in the next section.

\section{Action and variation}\label{sec:action}
The general form of the action for the dynamical fields listed in the previous section, for the class of theories we consider in this article, is given by
\begin{equation}\label{eqn:genaction}
S\left[\theta^a, \tp{\omega}^a{}_b, \phi, \chi^I\right] = S_g\left[\theta^a, \tp{\omega}^a{}_b, \phi\right] + S_m\left[\theta^a, \phi, \chi^I\right]\,,
\end{equation}
where \(S_g\) denotes the gravitational part of the action, while \(S_m\) denotes its matter part. Note in particular that we allow for a non-minimal coupling between the scalar field \(\phi\) and the matter fields \(\chi^I\), but no such couplings for the teleparallel spin connection.

Using the formalism of differential forms, the variation of the action with respect to the dynamical fields can be written in a very compact form. For the matter part \(S_m\) of the action we write a general variation as
\begin{equation}\label{eqn:gmatactvar}
\delta S_m = \int_M\left(\Sigma_a \wedge \delta\theta^a + \Psi \wedge \delta\phi + \Omega_I \wedge \delta\chi^I\right)\,,
\end{equation}
where we introduced the energy-momentum three-forms \(\Sigma_a\), a four-form \(\Psi\) and $(4 - k_I)$-forms \(\Omega_I\)\footnote{More precisely, the introduced objects \(\Sigma_a, \Psi, \Omega_I\) are twisted forms, which means that their sign changes under changing the orientation of the spacetime manifold. However, this distinction will not be relevant for the purpose of this article, and so we omit it for simplicity.}. Writing the variation in this form implies that any integration by parts, which is necessary in order to eliminate derivatives of the variations of the dynamical fields, has already been performed. Note that we wrote \(\Psi \wedge \delta\phi\), even though \(\delta\phi\) is a scalar; this is done simply for consistency of the notation.

For the variation of the gravitational part \(S_g\) of the action, we introduce a similar notation
\begin{equation}\label{eqn:ggravactcvar}
\delta S_g = \int_M\left(\Delta_a \wedge \delta\theta^a + \frac{1}{2}\Xi_a{}^b \wedge \delta\tp{\omega}^a{}_b + \Phi \wedge \delta\phi\right)\,,
\end{equation}
with three-forms \(\Delta_a\) and \(\Xi_a{}^b\), as well as a four-form \(\Phi\). We choose \(\Xi_a{}^b\) to be antisymmetric, \(\Xi_{(ab)} = 0\), since any symmetric part would cancel when contracted with the variation of the (also antisymmetric) Lorentz spin connection.

Using the one-to-one correspondence between spin connections and their torsion discussed in section~\ref{sec:fields}, one may also substitute the teleparallel spin connection in the gravitational action with the teleparallel torsion, and write its variation in the form
\begin{equation}\label{eqn:ggravacttvar}
\delta S_g = \int_M\left(\Upsilon_a \wedge \delta\theta^a + \Pi_a \wedge \delta T^a + \Phi \wedge \delta\phi\right)\,,
\end{equation}
with three-forms \(\Upsilon_a\), two-forms \(\Pi_a\) and the same four-form \(\Phi\) as above. Both forms of the variation can easily be related using the definition~\eqref{eqn:torsion} of the torsion, from which one derives the variation
\begin{equation}
\delta T^a = \delta\tp{\DD}\theta^a = \dd\delta\theta^a + \delta\tp{\omega}^a{}_b \wedge \theta^b + \tp{\omega}^a{}_b \wedge \delta\theta^b = \tp{\DD}\delta\theta^a + \delta\tp{\omega}^a{}_b \wedge \theta^b\,.
\end{equation}
Substituting this expression in the variation~\eqref{eqn:ggravacttvar} then yields
\begin{equation}
\delta S_g = \int_M\left[\left(\Upsilon_a - \tp{\DD}\Pi_a\right) \wedge \delta\theta^a - \Pi_a \wedge \theta^b \wedge \delta\tp{\omega}^a{}_b + \Phi \wedge \delta\phi\right]\,.
\end{equation}
By comparison with the variation~\eqref{eqn:ggravactcvar} one finds the relations
\begin{equation}\label{eqn:tvartocvar}
\Delta_a = \Upsilon_a - \tp{\DD}\Pi_a\,, \quad \Xi^{ab} = -2\Pi^{[a} \wedge \theta^{b]}\,.
\end{equation}
Conversely, one can make use of the relation~\eqref{eqn:invtorsion} to express the variation of the spin connection through the variation of the torsion. Using a similar procedure as given above one obtains
\begin{equation}\label{eqn:cvartotvar}
\Pi^a = \frac{1}{4}\iota_{e_c}\iota_{e_b}\Xi^{bc} \wedge \theta^a - \iota_{e_b}\Xi^{ab}\,, \quad \Upsilon^a = \Delta^a + \tp{\DD}\left(\frac{1}{4}\iota_{e_c}\iota_{e_b}\Xi^{bc} \wedge \theta^a - \iota_{e_b}\Xi^{ab}\right)\,.
\end{equation}
It is straightforward to check that the relations~\eqref{eqn:tvartocvar} and~\eqref{eqn:cvartotvar} are indeed inverses of each other.

In the following sections we will make use of these formulas, and consider the particular variations induced by local Lorentz transformations and diffeomorphisms.

\section{Local Lorentz invariance}\label{sec:loclor}
We now further demand that the action~\eqref{eqn:genaction} is invariant under (infinitesimal) local Lorentz transformations \(\lambda^a{}_b\) with \(\lambda^{(ab)} = 0\), which act on the tetrad and spin connection as
\begin{equation}
\delta_{\lambda}\theta^a = \lambda^a{}_b\theta^b\,, \quad \delta_{\lambda}\tp{\omega}^a{}_b = \lambda^a{}_c\tp{\omega}^c{}_b - \tp{\omega}^a{}_c\lambda^c{}_b - \dd\lambda^a{}_b = -\tp{\DD}\lambda^a{}_b\,.
\end{equation}
It thus follows that the torsion transforms as
\begin{equation}
\delta_{\lambda}T^a = \lambda^a{}_bT^b\,.
\end{equation}
We remark that the scalar field and the matter fields we consider are Lorentz scalars and hence transform trivially under local Lorentz transformations, such that \(\delta_{\lambda}\phi = 0\) and \(\delta_{\lambda}\chi^I = 0\).

We start with the variation of the matter action under Lorentz transformations, which is given by
\begin{equation}
\delta_{\lambda}S_m = \int_M\Sigma_a \wedge (\lambda^a{}_b\theta^b) = \int_M\Sigma^{[a} \wedge \theta^{b]}\lambda_{ab}\,.
\end{equation}
This vanishes for arbitrary local Lorentz transformations if and only if the energy-momentum 3-forms satisfy the symmetry condition
\begin{equation}\label{eqn:lorinvmat}
\Sigma^{[a} \wedge \theta^{b]} = 0\,.
\end{equation}
Note that this holds both on-shell and off-shell, since we have not used any field equations.

We then proceed analogously with the gravitational action. Writing its variation in the form~\eqref{eqn:ggravacttvar}, one finds that under local Lorentz transformations it transforms by
\begin{equation}
\delta_{\lambda}S_g = \int_M\left[\Upsilon_a \wedge (\lambda^a{}_b\theta^b) + \Pi_a \wedge \left(\lambda^a{}_bT^b\right)\right] = \int_M\left(\Upsilon^{[a} \wedge \theta^{b]} + \Pi^{[a} \wedge T^{b]}\right)\lambda_{ab}\,.
\end{equation}
It follows that the action is locally Lorentz invariant if and only if
\begin{equation}\label{eqn:lorinvgravt}
\Upsilon^{[a} \wedge \theta^{b]} + \Pi^{[a} \wedge T^{b]} = 0\,.
\end{equation}
Again we remark that this must hold both on-shell and off-shell. We can equivalently start from the variation~\eqref{eqn:ggravactcvar} and find
\begin{equation}
\delta_{\lambda}S_g = \int_M\left[\Delta_a \wedge (\lambda^a{}_b\theta^b) - \frac{1}{2}\Xi_a{}^b \wedge \tp{\DD}\lambda^a{}_b\right] = \int_M\left(\Delta^{[a} \wedge \theta^{b]} - \frac{1}{2}\tp{\DD}\Xi^{ab}\right)\lambda_{ab}\,.
\end{equation}
Hence, the condition for local Lorentz invariance reads
\begin{equation}\label{eqn:lorinvgravc}
\Delta^{[a} \wedge \theta^{b]} - \frac{1}{2}\tp{\DD}\Xi^{ab} = 0\,.
\end{equation}
Using the relations~\eqref{eqn:tvartocvar} and~\eqref{eqn:cvartotvar} one easily checks that the conditions~\eqref{eqn:lorinvgravt} and~\eqref{eqn:lorinvgravc} are equivalent.

\section{Diffeomorphism invariance and energy-momentum conservation}\label{sec:diffeo}
We now come to the discussion of diffeomorphism invariance. Recall that under an infinitesimal diffeomorphism generated by a vector field \(\xi\) any tensor field changes by its Lie derivative. For a differential form \(\tau\), the Lie derivative can be expressed as
\begin{equation}
\mathcal{L}_{\xi}\tau = \iota_{\xi}\dd\tau + \dd\iota_{\xi}\tau\,,
\end{equation}
which is also known as Cartan's (magic) formula.

We are in particular interested in the variation of the matter action under diffeomorphisms, which is given by
\begin{equation}
\delta_{\xi}S_m = \int_M\left(\Sigma_a \wedge \mathcal{L}_{\xi}\theta^a + \Psi \wedge \mathcal{L}_{\xi}\phi + \Omega_I \wedge \mathcal{L}_{\xi}\chi^I\right)\,.
\end{equation}
Note that on-shell the Euler-Lagrange equations \(\Omega_I = 0\) hold, so that the last term vanishes. We will therefore consider only the first two terms, keeping in mind that the resulting formulas hold only on-shell. Using the Cartan formula we can write
\begin{equation}
\begin{split}
\delta_{\xi}S_m &= \int_M\left(\Sigma_a \wedge \mathcal{L}_{\xi}\theta^a + \Psi \wedge \mathcal{L}_{\xi}\phi\right)\\
&= \int_M\left[\Sigma_a \wedge \left(\dd\iota_{\xi}\theta^a + \iota_{\xi}\dd\theta^a\right) + \Psi \wedge \iota_{\xi}\dd\phi\right]\\
&= \int_M\left(\dd\Sigma_a \wedge \iota_{\xi}\theta^a + \Sigma_a \wedge \iota_{\xi}\dd\theta^a + \Psi \wedge \iota_{\xi}\dd\phi\right)\\
&= \int_M\left(\dd\Sigma_a + \Sigma_b \wedge \iota_{e_a}\dd\theta^b + \Psi \wedge \iota_{e_a}\dd\phi\right)\xi^a\,.
\end{split}
\end{equation}
Here we have expressed \(\xi\) in the tetrad basis \(e_a\) to obtain the scalar functions \(\xi^a = \iota_{\xi}\theta^a\). The induced variation of the matter action vanishes for arbitrary vector fields \(\xi\) if and only if
\begin{equation}
\dd\Sigma_a + \Sigma_b \wedge \iota_{e_a}\dd\theta^b = -\Psi \wedge \iota_{e_a}\dd\phi\,.
\end{equation}
Note that for the Levi-Civita connection~\eqref{eqn:levicivita} holds
\begin{equation}
\begin{split}
\Sigma_b \wedge \lc{\omega}^b{}_a &= -\frac{1}{2}\left(\Sigma^b \wedge \theta^c\right)\left(\iota_{e_a}\iota_{e_c}\dd\theta_b + \iota_{e_c}\iota_{e_b}\dd\theta_a - \iota_{e_b}\iota_{e_a}\dd\theta_c\right)\\
&= -\left(\Sigma^b \wedge \theta^c\right)\iota_{e_a}\iota_{e_c}\dd\theta_b\\
&= \Sigma_b \wedge \iota_{e_a}\dd\theta^b\,.
\end{split}
\end{equation}
This finally allows us to write the diffeomorphism invariance condition as
\begin{equation}\label{eqn:enmomcons}
\dd\Sigma_a - \lc{\omega}^b{}_a \wedge \Sigma_b = \lc{\DD}\Sigma_a = -\Psi \wedge \iota_{e_a}\dd\phi\,.
\end{equation}
This is the covariant conservation of the energy-momentum 3-forms. In the case of a minimal (vanishing) coupling of the scalar field to the matter fields, \(\Psi = 0\), one obtains the usual energy-momentum conservation~\(\lc{\DD}\Sigma_a = 0\).

\section{Field equations}\label{sec:feqs}
We finally discuss the field equations of the general action~\eqref{eqn:genaction}. Already in the previous section we have mentioned the matter field equations, which take the simple form
\begin{equation}\label{eqn:feqmat}
\Omega_I = 0\,.
\end{equation}
The scalar field equation simply reads
\begin{equation}\label{eqn:feqscal}
\Phi + \Psi = 0\,.
\end{equation}
Next, we come to the tetrad field equations. These can most easily be read off from the variation~\eqref{eqn:ggravactcvar}, using the fact that the tetrad and the spin connection are varied independently, and take the form
\begin{equation}\label{eqn:feqtetc}
\Delta_a + \Sigma_a = 0\,.
\end{equation}
Using the relation~\eqref{eqn:tvartocvar}, they can also be written in the form
\begin{equation}\label{eqn:feqtett}
\Upsilon_a - \tp{\DD}\Pi_a + \Sigma_a = 0\,.
\end{equation}
Note that the structure of the tetrad field equations is the same as for the most general class of teleparallel gravity theories without a scalar field~\cite{Hohmann:2017duq}. The only difference lies in the fact that now all terms may carry an implicit dependence on the scalar field \(\phi\). It is helpful to consider the antisymmetric part of these field equations separately from the symmetric part. It can be written in different alternative forms. From the expressions~\eqref{eqn:feqtetc} and~\eqref{eqn:feqtett} we find
\begin{equation}\label{eqn:feqtetasym1}
\Delta^{[a} \wedge \theta^{b]} = \Upsilon^{[a} \wedge \theta^{b]} - \tp{\DD}\Pi^{[a} \wedge \theta^{b]} = 0\,,
\end{equation}
where we used the symmetry~\eqref{eqn:lorinvmat} of the energy-momentum three-forms. Using the Lorentz invariance conditions~\eqref{eqn:lorinvgravt} and~\eqref{eqn:lorinvgravc} we can also write
\begin{equation}\label{eqn:feqtetasym2}
\frac{1}{2}\tp{\DD}\Xi^{ab} = -\tp{\DD}\left(\Pi^{[a} \wedge \theta^{b]}\right) = -\tp{\DD}\Pi^{[a} \wedge \theta^{b]} - \Pi^{[a} \wedge T^{b]} = 0\,.
\end{equation}
Finally, we come to the field equation for the spin connection. Recall that the spin connection in teleparallel gravity theories is demanded to be flat, i.e., have vanishing curvature~\eqref{eqn:zerocurv}. Hence, we may allow only variations of the connection which preserve this condition, and must thus be of the form \(\delta\tp{\omega}^a{}_b = \tp{\DD}\pi^a{}_b\) with \(\pi_{(ab)} = 0\). From the variation~\eqref{eqn:ggravactcvar} one then immediately finds the field equations
\begin{equation}\label{eqn:feqconn}
\tp{\DD}\Xi^{ab} = 0\,.
\end{equation}
Note that these are simply the antisymmetric tetrad field equations~\eqref{eqn:feqtetasym2}, which follows from the fact that the teleparallel connection is a pure gauge degree of freedom. This is a direct generalization of the similar result for a number of sub-classes of teleparallel gravity theories~\cite{Golovnev:2017dox,Hohmann:2017duq,Hohmann:2018rwf} and one of the main results of this article.

\section{Conformal transformations}\label{sec:conf}
An interesting feature of the class of scalar-torsion theories discussed in this article is its invariance under conformal transformations of the tetrad and redefinitions of the scalar field. In order to show this, we consider a transformation of the tetrad and the scalar field of the form
\begin{equation}\label{eqn:conftrans}
\bar{\theta}^a = e^{\gamma(\phi)}\theta^a\,, \quad \bar{e}_a = e^{-\gamma(\phi)}e_a\,, \quad \bar{\phi} = f(\phi)\,,
\end{equation}
where \(\gamma\) and \(f\) are functions of the scalar field which determine the transformation. Note that we do not transform the spin connection, as it is a pure gauge degree of freedom, and any allowed transformation could be absorbed by a local Lorentz transformation. It then follows that the torsion transforms as
\begin{equation}\label{eqn:conftors}
\bar{T}^a = e^{\gamma}\left(T^a - \gamma'\theta^a \wedge \dd\phi\right)\,.
\end{equation}
In order to transform the terms in the field equations detailed in section~\ref{sec:feqs}, which have been obtained from the variations of the action displayed in section~\ref{sec:action}, we further need to transform the variations of the dynamical fields. For the variation of the tetrad one finds
\begin{equation}\label{eqn:confvartetrad}
\delta\bar{\theta}^a = e^{\gamma}\left(\delta\theta^a + \gamma'\theta^a\delta\phi\right)\,,
\end{equation}
while the variation of the scalar field transforms as
\begin{equation}\label{eqn:confvarscal}
\delta\bar{\phi} = f'\delta\phi\,.
\end{equation}
We also note that the variation of the torsion transforms as
\begin{equation}\label{eqn:confvartors}
\delta\bar{T}^a = e^{\gamma}\left[\delta T^a - \gamma'\delta\theta^a \wedge \dd\phi - \gamma'\theta^a \wedge \dd\delta\phi + \left(\gamma'T^a - \gamma'^2\theta^a \wedge \dd\phi - \gamma''\theta^a \wedge \dd\phi\right)\delta\phi\right]\,.
\end{equation}
Using these expressions, it is now straightforward to transform the variation of the action. For the variation~\eqref{eqn:gmatactvar} of the matter action we find
\begin{equation}
\begin{split}
\delta S_m &= \int_M\left(\bar{\Sigma}_a \wedge \delta\bar{\theta}^a + \bar{\Psi} \wedge \delta\bar{\phi} + \bar{\Omega}_I \wedge \delta\bar{\chi}^I\right)\\
&= \int_M\left[e^{\gamma}\bar{\Sigma}_a \wedge \delta\theta^a + \left(f'\bar{\Psi} + \gamma'e^{\gamma}\bar{\Sigma}_a \wedge \theta^a\right) \wedge \delta\phi + \bar{\Omega}_I \wedge \delta\chi^I\right]\,,
\end{split}
\end{equation}
By comparison with the original expression~\eqref{eqn:gmatactvar} we can thus read off the transformation rules
\begin{equation}
\Sigma_a = e^{\gamma}\bar{\Sigma}_a\,, \quad \Psi = f'\bar{\Psi} + \gamma'e^{\gamma}\bar{\Sigma}_a \wedge \theta^a\,, \quad \Omega_I = \bar{\Omega}_I\,.
\end{equation}
We finally invert these relations, in order to obtain the transformed (barred) quantities. These read
\begin{equation}\label{eqn:confvarm}
\bar{\Sigma}_a = e^{-\gamma}\Sigma_a\,, \quad \bar{\Psi} = \frac{1}{f'}\left(\Psi - \gamma'\Sigma_a \wedge \theta^a\right)\,, \quad \bar{\Omega}_I = \Omega_I\,.
\end{equation}
The same procedure can be applied to the gravitational part of the action. From the variation~\eqref{eqn:ggravactcvar} we obtain the transformation rules
\begin{equation}\label{eqn:confvargc}
\bar{\Delta}_a = e^{-\gamma}\Delta_a\,, \quad \bar{\Xi}_a{}^b = \Xi_a{}^b\,, \quad \bar{\Phi} = \frac{1}{f'}\left(\Phi - \gamma'\Delta_a \wedge \theta^a\right)\,.
\end{equation}
Applying the same procedure to the equivalent variation~\eqref{eqn:ggravacttvar} yields, after a more lengthy calculation, the corresponding transformation rules
\begin{equation}\label{eqn:confvargt}
\bar{\Upsilon}_a = e^{-\gamma}\left(\Upsilon_a - \gamma'\Pi_a \wedge \dd\phi\right)\,, \quad \bar{\Pi}_a = e^{-\gamma}\Pi_a\,, \quad \bar{\Phi} = \frac{1}{f'}\left[\Phi - \gamma'\left(\Upsilon_a - \tp{\DD}\Pi_a\right) \wedge \theta^a\right]\,.
\end{equation}
The same set of transformation rules can also be obtained by applying the transformation~\eqref{eqn:confvargc} to the relations~\eqref{eqn:tvartocvar} and~\eqref{eqn:cvartotvar}. One now easily checks that all dynamical and constraint equations derived in sections~\ref{sec:loclor}, \ref{sec:diffeo} and~\ref{sec:feqs} retain their form under these transformations, i.e., under replacing all quantities with their transformed (barred) counterparts. Note that for the energy-momentum conservation equation~\eqref{eqn:enmomcons} this includes replacing the Levi-Civita connection with the transformed connection
\begin{equation}\label{eqn:conflccon}
\lc{\bar{\omega}} = \lc{\omega} + \gamma'\left(\eta_{ac}\iota_{e_b}\dd\phi - \eta_{bc}\iota_{e_a}\dd\phi\right) \wedge \theta^c\,,
\end{equation}
which is obtained by applying the transformation~\eqref{eqn:conftrans} to the definition~\eqref{eqn:levicivita}. This yields the conservation equation
\begin{equation}
0 = \lc{\bar{\DD}}\bar{\Sigma}_a + \bar{\Psi} \wedge \iota_{\bar{e}_a}\dd\bar{\phi} = e^{-\gamma}\left(\lc{\DD}\Sigma_a + \Psi \wedge \iota_{e_a}\dd\phi\right)\,,
\end{equation}
whose proof requires also the symmetry~\eqref{eqn:lorinvmat} of the energy-momentum three-forms, and is omitted here for brevity. For the Lorentz invariance conditions displayed in section~\ref{sec:loclor} and the field equations shown in section~\ref{sec:feqs} the same notion of invariance under conformal transformations and redefinitions of the scalar field follows immediately, and will thus likewise be omitted here. This shows that the class of theories considered in this section, as well as all results derived, are invariant under these transformations, which is another important result we present in this article.

\section{Generalization to multiple scalar fields}\label{sec:multi}
Instead of a single scalar field \(\phi\) we now consider a scalar field multiplet \(\boldsymbol{\phi} = (\phi^A, A = 1, \ldots, N)\) of \(N\) scalar fields. The action~\eqref{eqn:genaction} of the theory now depends on all scalar fields, and thus takes the form
\begin{equation}\label{eqn:multigenaction}
S\left[\theta^a, \tp{\omega}^a{}_b, \phi^A, \chi^I\right] = S_g\left[\theta^a, \tp{\omega}^a{}_b, \phi^A\right] + S_m\left[\theta^a, \phi^A, \chi^I\right]\,.
\end{equation}
This action must accordingly be varied with respect to all scalar fields, and thus the expressions for the variation generalize correspondingly. For the variation of the matter action~\eqref{eqn:gmatactvar} we now have the structure
\begin{equation}\label{eqn:multimatactvar}
\delta S_m = \int_M\left(\Sigma_a \wedge \delta\theta^a + \Psi_A \wedge \delta\phi^A + \Omega_I \wedge \delta\chi^I\right)\,,
\end{equation}
while the variation~\eqref{eqn:ggravactcvar} of the gravitational action in connection variables reads
\begin{equation}\label{eqn:multigravactcvar}
\delta S_g = \int_M\left(\Delta_a \wedge \delta\theta^a + \frac{1}{2}\Xi_a{}^b \wedge \delta\tp{\omega}^a{}_b + \Phi_A \wedge \delta\phi^A\right)\,,
\end{equation}
or, equivalently, the variation~\eqref{eqn:ggravacttvar} in torsion variables,
\begin{equation}\label{eqn:multigravacttvar}
\delta S_g = \int_M\left(\Upsilon_a \wedge \delta\theta^a + \Pi_a \wedge \delta T^a + \Phi_A \wedge \delta\phi^A\right)\,.
\end{equation}
The dynamical equations derived from these variations receive only minor modifications. Most notably, the energy-momentum conservation~\eqref{eqn:enmomcons} for the matter fields now reads
\begin{equation}\label{eqn:multienmomcons}
\lc{\DD}\Sigma_a = -\Psi_A \wedge \iota_{e_a}\dd\phi^A\,,
\end{equation}
and there are \(N\) scalar field equations of the form~\eqref{eqn:feqscal},
\begin{equation}\label{eqn:multifeqscal}
\Phi_A + \Psi_A = 0\,.
\end{equation}
However, all other field equations retain their structure, although the terms therein depend on a larger number of scalar fields. Also the constraint equations associated to local Lorentz invariance detailed in section~\ref{sec:loclor} remain unchanged, since the scalar fields are not affected by local Lorentz transformations.

The difference between a single and multiple scalar fields becomes most apparent in the discussion of invariance under conformal transformations and scalar field redefinitions, which we presented in section~\ref{sec:conf}. In the multi-scalar-torsion case the combined transformation~\eqref{eqn:conftrans} reads
\begin{equation}\label{eqn:multiconftrans}
\bar{\theta}^a = e^{\gamma(\boldsymbol{\phi})}\theta^a\,, \quad \bar{e}_a = e^{-\gamma(\boldsymbol{\phi})}e_a\,, \quad \bar{\phi}^A = f^A(\boldsymbol{\phi})\,.
\end{equation}
Hence, instead of a single function \(f\) there are \(N\) functions \(f^A\) defining the scalar field redefinition, and these functions as well as the conformal transformation \(\gamma\) depend on the multiplet \(\boldsymbol{\phi}\) of all scalar fields. Alternatively, we may write \(\bar{\boldsymbol{\phi}} = \boldsymbol{f}(\boldsymbol{\phi})\) for the scalar field redefinition. Also the formulas for the transformation behavior of derived quantities receive correction terms. In particular, we find the transformation~\eqref{eqn:conftors} of the torsion
\begin{equation}\label{eqn:multiconftors}
\bar{T}^a = e^{\gamma}\left(T^a - \gamma_{,A}\theta^a \wedge \dd\phi^A\right)\,,
\end{equation}
as well as the transformation~\eqref{eqn:conflccon} of the Levi-Civita connection
\begin{equation}
\lc{\bar{\omega}} = \lc{\omega} + \gamma_{,A}\left(\eta_{ac}\iota_{e_b}\dd\phi^A - \eta_{bc}\iota_{e_a}\dd\phi^A\right) \wedge \theta^c\,,
\end{equation}
where we introduced the notation
\begin{equation}
\gamma_{,A} = \frac{\partial}{\partial\phi^A}\gamma
\end{equation}
for the derivative with respect to one of the scalar fields. This now allows us to generalize the transformation behavior of the variation of the action. In particular, the formula~\eqref{eqn:confvartetrad} generalizes to
\begin{equation}
\delta\bar{\theta}^a = e^{\gamma}\left(\delta\theta^a + \gamma_{,A}\theta^a\delta\phi^A\right)\,,
\end{equation}
while the formula~\eqref{eqn:confvarscal} generalizes to
\begin{equation}
\delta\bar{\phi}^A = \frac{\partial\bar{\phi}^A}{\partial\phi^B}\delta\phi^B\,.
\end{equation}
Here the factor on the right hand side is simply the Jacobian of the function \(\boldsymbol{f}\), which determines the scalar field redefinition. Further, the variation of the torsion transforms as
\begin{equation}
\delta\bar{T}^a = e^{\gamma}\left[\delta T^a - \gamma_{,A}\delta\theta^a \wedge \dd\phi^A - \gamma_{,A}\theta^a \wedge \dd\delta\phi^A + \left(\gamma_{,A}T^a - \gamma_{,A}\gamma_{,B}\theta^a \wedge \dd\phi^B - \gamma_{,AB}\theta^a \wedge \dd\phi^B\right)\delta\phi^A\right]\,,
\end{equation}
generalizing the transformation formula~\eqref{eqn:confvartors}. By comparison with the variation of the action we then find that the terms in the field equations transform as
\begin{equation}
\bar{\Psi}_A = \frac{\partial\phi^B}{\partial\bar{\phi}^A}\left(\Psi_B - \gamma_{,B}\Sigma_a \wedge \theta^a\right)\,, \quad \bar{\Phi}_A = \frac{\partial\phi^B}{\partial\bar{\phi}^A}\left(\Phi_B - \gamma_{,B}\Delta_a \wedge \theta^a\right)\,, \quad \bar{\Upsilon}_a = e^{-\gamma}\left(\Upsilon_a - \gamma_{,A}\Pi_a \wedge \dd\phi^A\right)\,,
\end{equation}
while the remaining terms keep their transformation behavior detailed by the relations~\eqref{eqn:confvarm}, \eqref{eqn:confvargc} and~\eqref{eqn:confvargt}. One now easily checks that also this larger class of theories, with \(N\) scalar fields instead of a single field, satisfies the same notion of being closed under conformal transformations of the type~\eqref{eqn:multiconftrans}, in the sense that the action and the field equations retain their form.

\section{Examples}\label{sec:examples}
In order to show the applicability of the results we obtained in the previous sections, we finally display two examples. Here we restrict ourselves to classes of theories, which are by themselves very general, and contain a wide range of well studied teleparallel scalar-torsion models. In particular, we discuss gravity theories defined by a constitutive relation in section~\ref{ssec:constrel}, and a newly introduced class of theories defined by a free function of four scalar quantities in section~\ref{ssec:ltxyp}.

\subsection{Constitutive relations and second order field equations}\label{ssec:constrel}
In a recent work~\cite{Hohmann:2017duq} we showed how the action of any teleparallel gravity theory with second order field equations can be expressed in the form
\begin{equation}
S_g\left[\theta^a, \tp{\omega}^a{}_b\right] = \frac{1}{2}\int_MT^a \wedge H_a\left(\theta^a, T^a\right)\,,
\end{equation}
where the excitation two-forms \(H_a\) are defined by a constitutive relation as an algebraic function \(H_a(\theta^a, T^a)\) of the coframe and the torsion. We argued that also a scalar field may be included in the constitutive relation, \(H_a(\theta^a, T^a, \phi)\), and gave a specific example. This motivates considering a class of theories based on a constitutive relation involving an arbitrary number of scalar fields, whose action reads
\begin{equation}\label{eqn:consaction}
S_g\left[\theta^a, \tp{\omega}^a{}_b, \phi^A\right] = \frac{1}{2}\int_MT^a \wedge H_a\left(\theta^a, T^a, \phi^A\right) + S_{\phi}\left[\theta^a, \phi^A\right]\,,
\end{equation}
where the scalar field action \(S_{\phi}\) determines the dynamics of the scalar fields through suitable kinetic and potential terms, and is also assumed to be locally Lorentz invariant. It is now straightforward to apply the analysis performed in this article to this class of theories; see~\cite{Hohmann:2017duq}, where this is already partially done.

We conclude the discussion of this class of theories with a remark on its behavior under conformal transformations. It can be seen from the transformation laws~\eqref{eqn:conftors} and~\eqref{eqn:multiconftors} of the torsion that the appearing factor \(e^{\gamma}\), which rescales the torsion tensor, can be absorbed into the transformation of the constitutive relation \(H_a(\theta^a, T^a, \phi)\). However, this is not the case for the additive term involving derivatives of the scalar field, which is in general non-minimally coupled to torsion, and thus cannot be decomposed into contributions to the constitutive relation and the scalar field action \(S_{\phi}\).

\subsection{$L(T, X, Y, \phi)$ theory}\label{ssec:ltxyp}
The non-minimal derivative coupling between the scalar field and the torsion tensor arising from a conformal transformation of the action~\eqref{eqn:consaction} motivates studying actions in which such terms are included. We therefore propose a new generic class of theories, which is constructed by assuming a Lagrangian that is a free function of four scalar quantities, which we define as follows. First, we write the torsion in the tetrad basis as
\begin{equation}
T^a = \frac{1}{2}T^a{}_{bc}\theta^b \wedge \theta^c\,, \quad T^a{}_{bc} = \iota_{e_c}\iota_{e_b}T^a\,.
\end{equation}
We then define the superpotential
\begin{equation}
S_a{}^{bc} = \frac{1}{2}\chi_a{}^{bc}{}_d{}^{ef}T^d{}_{ef}\,,
\end{equation}
where the constitutive tensor is given by
\begin{equation}
\chi_a{}^{bc}{}_d{}^{ef} = \eta_{ad}\eta^{e[b}\eta^{c]f} + 2\delta_d^{[c}\eta^{b][e}\delta_a^{f]} - 4\delta_a^{[c}\eta^{b][e}\delta_d^{f]}\,.
\end{equation}
From this we can define the torsion scalar
\begin{equation}
T = \frac{1}{2}T^a{}_{bc}S_a{}^{bc} = \frac{1}{4}T^{abc}T_{abc} + \frac{1}{2}T^{abc}T_{cba} - T^a{}_{ba}T^{cb}{}_c\,,
\end{equation}
We further define the scalar field kinetic term
\begin{equation}
X = -\frac{1}{2}\eta^{ab}\iota_{e_a}\dd\phi\iota_{e_b}\dd\phi\,,
\end{equation}
as well as the kinetic coupling term
\begin{equation}
Y = T^a{}_{ab}\iota_{e_c}\dd\phi\eta^{bc}\,.
\end{equation}
Finally, the gravitational Lagrangian of the theory if given as a free function of the three aforementioned quantities and the scalar field, such that the gravitational part of the action reads
\begin{equation}\label{eqn:confaction}
S_g\left[\theta^a, \tp{\omega}^a{}_b, \phi^A\right] = \int_ML(T, X, Y, \phi)\,\theta^0 \wedge \theta^1 \wedge \theta^2 \wedge \theta^3\,.
\end{equation}
With the definitions given above it is now straightforward to use the formalism detailed in this article in order to derive the gravitational field equations; see~\cite{Hohmann:2018dqh} for a detailed discussion and full derivation of various properties of this class of theories. In particular, we show that this class of theories is algebraically closed under conformal transformation, i.e., a conformal transformation of the action~\eqref{eqn:confaction} yields another action of the same form, but with a different functional form of the Lagrangian \(L\). We finally remark that also in this case the field equations are (at most) of second derivative order, since the action functional contains at most first order derivatives of the fundamental field variables.

\section{Conclusion}\label{sec:conclusion}
In this article we have discussed the most general class of scalar-torsion theories of gravity, in which one or more scalar fields are coupled to the tetrad and flat spin connection of teleparallel geometry. We have restricted the theories only by demanding invariance under local Lorentz transformations and diffeomorphisms, as well as by excluding a direct coupling of matter fields to the teleparallel spin connection. As the main result, we have investigated a number of statements, which have been previously discussed in the context of narrower classes of teleparallel gravitational theories, and proven that they also hold for this very broad class of theories. In particular, we have shown how the condition of local Lorentz invariance causes the field equations for the teleparallel spin connection to be equal to the antisymmetric part of the tetrad field equations. As a consequence, the spin connection becomes a pure gauge degree of freedom. Further, we have shown that conformal transformations of the tetrad and scalar field redefinitions relate different theories within this class.

In order to show the applicability of the results we derived in this work, we listed two classes of example theories which are covered by the formalism we derived, and which by itself are broad enough to contain plenty of interesting theories and subclasses. Both classes have in common that their field equations are at most of second derivative order, since their actions are constructed only from terms that are of first derivative order in the fundamental fields. However, note that we made no restriction on the derivative orders in the theories we considered in this article. Hence, more general examples, involving also higher derivatives, may be conceived.

Various other aspects of the generic class of scalar-tensor theories we discussed here may be studied. For example, it may be discussed how spacetime symmetries can be exploited in order to narrow the possible solutions for the fundamental fields, to simplify the field equations, and potentially derive generic solutions for a larger class of theories~\cite{Hohmann:2015pva,Hohmann:2018sym}. One may also consider using the derivations presented here in order to discuss phenomenological aspects, or construct a general formalism for the phenomenology of scalar-torsion theories of gravity. Finally, one may exploit alternative formulations of teleparallel gravity, for example, as a higher gauge theory~\cite{Baez:2012bn}, in order to extend our results to related classes of gravity theories based on more general geometric backgrounds.

The work we presented here is foundational, and allows for various generalizations. Here we considered only scalar fields which are non-minimally coupled to the tetrad and spin connection of teleparallel geometry. One may relax this condition and also allow for more general tensor fields in the gravitational part of the action. Of particular interest could be additional fields carrying Lorentz indices, which would therefore transform non-trivially under local Lorentz transformations. As with the scalar fields in this article, one may also allow these additional fields to couple non-minimally to matter fields. Finally, one may relax the condition that there is no direct coupling between matter and the teleparallel spin connection. This would be interesting in particular in the presence of spinning matter, which is the source of torsion in Einstein-Cartan gravity, where torsion appears as an independent degree of freedom in addition to the metric.

\begin{acknowledgments}
The author thanks Martin Kr\v{s}\v{s}\'ak and Christian Pfeifer for helpful comments and discussions. He gratefully acknowledges the full financial support of the Estonian Ministry for Education and Science through the Institutional Research Support Project IUT02-27 and Startup Research Grant PUT790, as well as the European Regional Development Fund through the Center of Excellence TK133 ``The Dark Side of the Universe''.
\end{acknowledgments}

\bibliography{scaltors}

\begin{thebibliography}{31}%
\makeatletter
\providecommand \@ifxundefined [1]{%
 \@ifx{#1\undefined}
}%
\providecommand \@ifnum [1]{%
 \ifnum #1\expandafter \@firstoftwo
 \else \expandafter \@secondoftwo
 \fi
}%
\providecommand \@ifx [1]{%
 \ifx #1\expandafter \@firstoftwo
 \else \expandafter \@secondoftwo
 \fi
}%
\providecommand \natexlab [1]{#1}%
\providecommand \enquote  [1]{``#1''}%
\providecommand \bibnamefont  [1]{#1}%
\providecommand \bibfnamefont [1]{#1}%
\providecommand \citenamefont [1]{#1}%
\providecommand \href@noop [0]{\@secondoftwo}%
\providecommand \href [0]{\begingroup \@sanitize@url \@href}%
\providecommand \@href[1]{\@@startlink{#1}\@@href}%
\providecommand \@@href[1]{\endgroup#1\@@endlink}%
\providecommand \@sanitize@url [0]{\catcode `\\12\catcode `\$12\catcode
  `\&12\catcode `\#12\catcode `\^12\catcode `\_12\catcode `\%12\relax}%
\providecommand \@@startlink[1]{}%
\providecommand \@@endlink[0]{}%
\providecommand \url  [0]{\begingroup\@sanitize@url \@url }%
\providecommand \@url [1]{\endgroup\@href {#1}{\urlprefix }}%
\providecommand \urlprefix  [0]{URL }%
\providecommand \Eprint [0]{\href }%
\providecommand \doibase [0]{http://dx.doi.org/}%
\providecommand \selectlanguage [0]{\@gobble}%
\providecommand \bibinfo  [0]{\@secondoftwo}%
\providecommand \bibfield  [0]{\@secondoftwo}%
\providecommand \translation [1]{[#1]}%
\providecommand \BibitemOpen [0]{}%
\providecommand \bibitemStop [0]{}%
\providecommand \bibitemNoStop [0]{.\EOS\space}%
\providecommand \EOS [0]{\spacefactor3000\relax}%
\providecommand \BibitemShut  [1]{\csname bibitem#1\endcsname}%
\let\auto@bib@innerbib\@empty
\bibitem [{\citenamefont {Ade}\ \emph {et~al.}(2016{\natexlab{a}})\citenamefont
  {Ade} \emph {et~al.}}]{Ade:2015xua}%
  \BibitemOpen
  \bibfield  {author} {\bibinfo {author} {\bibfnamefont {P.~A.~R.}\
  \bibnamefont {Ade}} \emph {et~al.} (\bibinfo {collaboration} {Planck}),\
  }\bibfield  {title} {\enquote {\bibinfo {title} {{Planck 2015 results. XIII.
  Cosmological parameters}},}\ }\href {\doibase 10.1051/0004-6361/201525830}
  {\bibfield  {journal} {\bibinfo  {journal} {Astron. Astrophys.}\ }\textbf
  {\bibinfo {volume} {594}},\ \bibinfo {pages} {A13} (\bibinfo {year}
  {2016}{\natexlab{a}})},\ \Eprint {http://arxiv.org/abs/1502.01589}
  {arXiv:1502.01589 [astro-ph.CO]} \BibitemShut {NoStop}%
\bibitem [{\citenamefont {Ade}\ \emph {et~al.}(2016{\natexlab{b}})\citenamefont
  {Ade} \emph {et~al.}}]{Ade:2015rim}%
  \BibitemOpen
  \bibfield  {author} {\bibinfo {author} {\bibfnamefont {P.~A.~R.}\
  \bibnamefont {Ade}} \emph {et~al.} (\bibinfo {collaboration} {Planck}),\
  }\bibfield  {title} {\enquote {\bibinfo {title} {{Planck 2015 results. XIV.
  Dark energy and modified gravity}},}\ }\href {\doibase
  10.1051/0004-6361/201525814} {\bibfield  {journal} {\bibinfo  {journal}
  {Astron. Astrophys.}\ }\textbf {\bibinfo {volume} {594}},\ \bibinfo {pages}
  {A14} (\bibinfo {year} {2016}{\natexlab{b}})},\ \Eprint
  {http://arxiv.org/abs/1502.01590} {arXiv:1502.01590 [astro-ph.CO]}
  \BibitemShut {NoStop}%
\bibitem [{\citenamefont {Ade}\ \emph {et~al.}(2016{\natexlab{c}})\citenamefont
  {Ade} \emph {et~al.}}]{Ade:2015lrj}%
  \BibitemOpen
  \bibfield  {author} {\bibinfo {author} {\bibfnamefont {P.~A.~R.}\
  \bibnamefont {Ade}} \emph {et~al.} (\bibinfo {collaboration} {Planck}),\
  }\bibfield  {title} {\enquote {\bibinfo {title} {{Planck 2015 results. XX.
  Constraints on inflation}},}\ }\href {\doibase 10.1051/0004-6361/201525898}
  {\bibfield  {journal} {\bibinfo  {journal} {Astron. Astrophys.}\ }\textbf
  {\bibinfo {volume} {594}},\ \bibinfo {pages} {A20} (\bibinfo {year}
  {2016}{\natexlab{c}})},\ \Eprint {http://arxiv.org/abs/1502.02114}
  {arXiv:1502.02114 [astro-ph.CO]} \BibitemShut {NoStop}%
\bibitem [{\citenamefont {M{\o}ller}(1961)}]{Moller:1961}%
  \BibitemOpen
  \bibfield  {author} {\bibinfo {author} {\bibfnamefont {Christian}\
  \bibnamefont {M{\o}ller}},\ }\bibfield  {title} {\enquote {\bibinfo {title}
  {{Conservation Laws and Absolute Parallelism in General Relativity}},}\
  }\href@noop {} {\bibfield  {journal} {\bibinfo  {journal} {K. Dan. Vidensk.
  Selsk. Mat. Fys. Skr.}\ }\textbf {\bibinfo {volume} {1}},\ \bibinfo {pages}
  {1--50} (\bibinfo {year} {1961})}\BibitemShut {NoStop}%
\bibitem [{\citenamefont {Aldrovandi}\ and\ \citenamefont
  {Pereira}(2013)}]{Aldrovandi:2013wha}%
  \BibitemOpen
  \bibfield  {author} {\bibinfo {author} {\bibfnamefont {Ruben}\ \bibnamefont
  {Aldrovandi}}\ and\ \bibinfo {author} {\bibfnamefont {José~Geraldo}\
  \bibnamefont {Pereira}},\ }\href {\doibase 10.1007/978-94-007-5143-9} {\emph
  {\bibinfo {title} {{Teleparallel Gravity}}}},\ Vol.\ \bibinfo {volume} {173}\
  (\bibinfo  {publisher} {Springer},\ \bibinfo {address} {Dordrecht},\ \bibinfo
  {year} {2013})\BibitemShut {NoStop}%
\bibitem [{\citenamefont {Maluf}(2013)}]{Maluf:2013gaa}%
  \BibitemOpen
  \bibfield  {author} {\bibinfo {author} {\bibfnamefont {J.~W.}\ \bibnamefont
  {Maluf}},\ }\bibfield  {title} {\enquote {\bibinfo {title} {{The teleparallel
  equivalent of general relativity}},}\ }\href {\doibase
  10.1002/andp.201200272} {\bibfield  {journal} {\bibinfo  {journal} {Annalen
  Phys.}\ }\textbf {\bibinfo {volume} {525}},\ \bibinfo {pages} {339--357}
  (\bibinfo {year} {2013})},\ \Eprint {http://arxiv.org/abs/1303.3897}
  {arXiv:1303.3897 [gr-qc]} \BibitemShut {NoStop}%
\bibitem [{\citenamefont {Golovnev}(2018)}]{Golovnev:2018red}%
  \BibitemOpen
  \bibfield  {author} {\bibinfo {author} {\bibfnamefont {Alexey}\ \bibnamefont
  {Golovnev}},\ }\bibfield  {title} {\enquote {\bibinfo {title} {{Introduction
  to teleparallel gravities}},}\ }in\ \href
  {http://inspirehep.net/record/1649207/files/arXiv:1801.06929.pdf} {\emph
  {\bibinfo {booktitle} {{9th Mathematical Physics Meeting: Summer School and
  Conference on Modern Mathematical Physics Belgrade, Serbia, September 18-23,
  2017}}}}\ (\bibinfo {year} {2018})\ \Eprint {http://arxiv.org/abs/1801.06929}
  {arXiv:1801.06929 [gr-qc]} \BibitemShut {NoStop}%
\bibitem [{\citenamefont {Cho}(1976{\natexlab{a}})}]{Cho:1975dh}%
  \BibitemOpen
  \bibfield  {author} {\bibinfo {author} {\bibfnamefont {Y.~M.}\ \bibnamefont
  {Cho}},\ }\bibfield  {title} {\enquote {\bibinfo {title} {{Einstein
  Lagrangian as the Translational Yang-Mills Lagrangian}},}\ }\href {\doibase
  10.1103/PhysRevD.14.2521} {\bibfield  {journal} {\bibinfo  {journal} {Phys.
  Rev.}\ }\textbf {\bibinfo {volume} {D14}},\ \bibinfo {pages} {2521} (\bibinfo
  {year} {1976}{\natexlab{a}})}\BibitemShut {NoStop}%
\bibitem [{\citenamefont {Cho}(1976{\natexlab{b}})}]{Cho:1976fr}%
  \BibitemOpen
  \bibfield  {author} {\bibinfo {author} {\bibfnamefont {Y.~M.}\ \bibnamefont
  {Cho}},\ }\bibfield  {title} {\enquote {\bibinfo {title} {{Gauge Theory of
  Poincare Symmetry}},}\ }\href {\doibase 10.1103/PhysRevD.14.3335} {\bibfield
  {journal} {\bibinfo  {journal} {Phys. Rev.}\ }\textbf {\bibinfo {volume}
  {D14}},\ \bibinfo {pages} {3335--3340} (\bibinfo {year}
  {1976}{\natexlab{b}})}\BibitemShut {NoStop}%
\bibitem [{\citenamefont {Hayashi}(1977)}]{Hayashi:1977jd}%
  \BibitemOpen
  \bibfield  {author} {\bibinfo {author} {\bibfnamefont {Kenji}\ \bibnamefont
  {Hayashi}},\ }\bibfield  {title} {\enquote {\bibinfo {title} {{The Gauge
  Theory of the Translation Group and Underlying Geometry}},}\ }\href {\doibase
  10.1016/0370-2693(77)90840-1} {\bibfield  {journal} {\bibinfo  {journal}
  {Phys. Lett.}\ }\textbf {\bibinfo {volume} {B69}},\ \bibinfo {pages}
  {441--444} (\bibinfo {year} {1977})}\BibitemShut {NoStop}%
\bibitem [{\citenamefont {Li}\ \emph {et~al.}(2011{\natexlab{a}})\citenamefont
  {Li}, \citenamefont {Sotiriou},\ and\ \citenamefont {Barrow}}]{Li:2010cg}%
  \BibitemOpen
  \bibfield  {author} {\bibinfo {author} {\bibfnamefont {Baojiu}\ \bibnamefont
  {Li}}, \bibinfo {author} {\bibfnamefont {Thomas~P.}\ \bibnamefont
  {Sotiriou}}, \ and\ \bibinfo {author} {\bibfnamefont {John~D.}\ \bibnamefont
  {Barrow}},\ }\bibfield  {title} {\enquote {\bibinfo {title} {{$f(T)$ gravity
  and local Lorentz invariance}},}\ }\href {\doibase
  10.1103/PhysRevD.83.064035} {\bibfield  {journal} {\bibinfo  {journal} {Phys.
  Rev.}\ }\textbf {\bibinfo {volume} {D83}},\ \bibinfo {pages} {064035}
  (\bibinfo {year} {2011}{\natexlab{a}})},\ \Eprint
  {http://arxiv.org/abs/1010.1041} {arXiv:1010.1041 [gr-qc]} \BibitemShut
  {NoStop}%
\bibitem [{\citenamefont {Sotiriou}\ \emph {et~al.}(2011)\citenamefont
  {Sotiriou}, \citenamefont {Li},\ and\ \citenamefont
  {Barrow}}]{Sotiriou:2010mv}%
  \BibitemOpen
  \bibfield  {author} {\bibinfo {author} {\bibfnamefont {Thomas~P.}\
  \bibnamefont {Sotiriou}}, \bibinfo {author} {\bibfnamefont {Baojiu}\
  \bibnamefont {Li}}, \ and\ \bibinfo {author} {\bibfnamefont {John~D.}\
  \bibnamefont {Barrow}},\ }\bibfield  {title} {\enquote {\bibinfo {title}
  {{Generalizations of teleparallel gravity and local Lorentz symmetry}},}\
  }\href {\doibase 10.1103/PhysRevD.83.104030} {\bibfield  {journal} {\bibinfo
  {journal} {Phys. Rev.}\ }\textbf {\bibinfo {volume} {D83}},\ \bibinfo {pages}
  {104030} (\bibinfo {year} {2011})},\ \Eprint {http://arxiv.org/abs/1012.4039}
  {arXiv:1012.4039 [gr-qc]} \BibitemShut {NoStop}%
\bibitem [{\citenamefont {Li}\ \emph {et~al.}(2011{\natexlab{b}})\citenamefont
  {Li}, \citenamefont {Miao},\ and\ \citenamefont {Miao}}]{Li:2011rn}%
  \BibitemOpen
  \bibfield  {author} {\bibinfo {author} {\bibfnamefont {Miao}\ \bibnamefont
  {Li}}, \bibinfo {author} {\bibfnamefont {Rong-Xin}\ \bibnamefont {Miao}}, \
  and\ \bibinfo {author} {\bibfnamefont {Yan-Gang}\ \bibnamefont {Miao}},\
  }\bibfield  {title} {\enquote {\bibinfo {title} {{Degrees of freedom of
  $f(T)$ gravity}},}\ }\href {\doibase 10.1007/JHEP07(2011)108} {\bibfield
  {journal} {\bibinfo  {journal} {JHEP}\ }\textbf {\bibinfo {volume} {07}},\
  \bibinfo {pages} {108} (\bibinfo {year} {2011}{\natexlab{b}})},\ \Eprint
  {http://arxiv.org/abs/1105.5934} {arXiv:1105.5934 [hep-th]} \BibitemShut
  {NoStop}%
\bibitem [{\citenamefont {Ong}\ \emph {et~al.}(2013)\citenamefont {Ong},
  \citenamefont {Izumi}, \citenamefont {Nester},\ and\ \citenamefont
  {Chen}}]{Ong:2013qja}%
  \BibitemOpen
  \bibfield  {author} {\bibinfo {author} {\bibfnamefont {Yen~Chin}\
  \bibnamefont {Ong}}, \bibinfo {author} {\bibfnamefont {Keisuke}\ \bibnamefont
  {Izumi}}, \bibinfo {author} {\bibfnamefont {James~M.}\ \bibnamefont
  {Nester}}, \ and\ \bibinfo {author} {\bibfnamefont {Pisin}\ \bibnamefont
  {Chen}},\ }\bibfield  {title} {\enquote {\bibinfo {title} {{Problems with
  Propagation and Time Evolution in f(T) Gravity}},}\ }\href {\doibase
  10.1103/PhysRevD.88.024019} {\bibfield  {journal} {\bibinfo  {journal} {Phys.
  Rev.}\ }\textbf {\bibinfo {volume} {D88}},\ \bibinfo {pages} {024019}
  (\bibinfo {year} {2013})},\ \Eprint {http://arxiv.org/abs/1303.0993}
  {arXiv:1303.0993 [gr-qc]} \BibitemShut {NoStop}%
\bibitem [{\citenamefont {Izumi}\ \emph {et~al.}(2014)\citenamefont {Izumi},
  \citenamefont {Gu},\ and\ \citenamefont {Ong}}]{Izumi:2013dca}%
  \BibitemOpen
  \bibfield  {author} {\bibinfo {author} {\bibfnamefont {Keisuke}\ \bibnamefont
  {Izumi}}, \bibinfo {author} {\bibfnamefont {Je-An}\ \bibnamefont {Gu}}, \
  and\ \bibinfo {author} {\bibfnamefont {Yen~Chin}\ \bibnamefont {Ong}},\
  }\bibfield  {title} {\enquote {\bibinfo {title} {{Acausality and Nonunique
  Evolution in Generalized Teleparallel Gravity}},}\ }\href {\doibase
  10.1103/PhysRevD.89.084025} {\bibfield  {journal} {\bibinfo  {journal} {Phys.
  Rev.}\ }\textbf {\bibinfo {volume} {D89}},\ \bibinfo {pages} {084025}
  (\bibinfo {year} {2014})},\ \Eprint {http://arxiv.org/abs/1309.6461}
  {arXiv:1309.6461 [gr-qc]} \BibitemShut {NoStop}%
\bibitem [{\citenamefont {Chen}\ \emph
  {et~al.}(2015{\natexlab{a}})\citenamefont {Chen}, \citenamefont {Izumi},
  \citenamefont {Nester},\ and\ \citenamefont {Ong}}]{Chen:2014qtl}%
  \BibitemOpen
  \bibfield  {author} {\bibinfo {author} {\bibfnamefont {Pisin}\ \bibnamefont
  {Chen}}, \bibinfo {author} {\bibfnamefont {Keisuke}\ \bibnamefont {Izumi}},
  \bibinfo {author} {\bibfnamefont {James~M.}\ \bibnamefont {Nester}}, \ and\
  \bibinfo {author} {\bibfnamefont {Yen~Chin}\ \bibnamefont {Ong}},\ }\bibfield
   {title} {\enquote {\bibinfo {title} {{Remnant Symmetry, Propagation and
  Evolution in $f$(T) Gravity}},}\ }\href {\doibase 10.1103/PhysRevD.91.064003}
  {\bibfield  {journal} {\bibinfo  {journal} {Phys. Rev.}\ }\textbf {\bibinfo
  {volume} {D91}},\ \bibinfo {pages} {064003} (\bibinfo {year}
  {2015}{\natexlab{a}})},\ \Eprint {http://arxiv.org/abs/1412.8383}
  {arXiv:1412.8383 [gr-qc]} \BibitemShut {NoStop}%
\bibitem [{\citenamefont {Krššák}\ and\ \citenamefont
  {Saridakis}(2016)}]{Krssak:2015oua}%
  \BibitemOpen
  \bibfield  {author} {\bibinfo {author} {\bibfnamefont {Martin}\ \bibnamefont
  {Krššák}}\ and\ \bibinfo {author} {\bibfnamefont {Emmanuel~N.}\
  \bibnamefont {Saridakis}},\ }\bibfield  {title} {\enquote {\bibinfo {title}
  {{The covariant formulation of f(T) gravity}},}\ }\href {\doibase
  10.1088/0264-9381/33/11/115009} {\bibfield  {journal} {\bibinfo  {journal}
  {Class. Quant. Grav.}\ }\textbf {\bibinfo {volume} {33}},\ \bibinfo {pages}
  {115009} (\bibinfo {year} {2016})},\ \Eprint
  {http://arxiv.org/abs/1510.08432} {arXiv:1510.08432 [gr-qc]} \BibitemShut
  {NoStop}%
\bibitem [{\citenamefont {Golovnev}\ \emph {et~al.}(2017)\citenamefont
  {Golovnev}, \citenamefont {Koivisto},\ and\ \citenamefont
  {Sandstad}}]{Golovnev:2017dox}%
  \BibitemOpen
  \bibfield  {author} {\bibinfo {author} {\bibfnamefont {Alexey}\ \bibnamefont
  {Golovnev}}, \bibinfo {author} {\bibfnamefont {Tomi}\ \bibnamefont
  {Koivisto}}, \ and\ \bibinfo {author} {\bibfnamefont {Marit}\ \bibnamefont
  {Sandstad}},\ }\bibfield  {title} {\enquote {\bibinfo {title} {{On the
  covariance of teleparallel gravity theories}},}\ }\href {\doibase
  10.1088/1361-6382/aa7830} {\bibfield  {journal} {\bibinfo  {journal} {Class.
  Quant. Grav.}\ }\textbf {\bibinfo {volume} {34}},\ \bibinfo {pages} {145013}
  (\bibinfo {year} {2017})},\ \Eprint {http://arxiv.org/abs/1701.06271}
  {arXiv:1701.06271 [gr-qc]} \BibitemShut {NoStop}%
\bibitem [{\citenamefont {Krssak}(2017)}]{Krssak:2017nlv}%
  \BibitemOpen
  \bibfield  {author} {\bibinfo {author} {\bibfnamefont {Martin}\ \bibnamefont
  {Krssak}},\ }\bibfield  {title} {\enquote {\bibinfo {title} {{Variational
  Problem and Bigravity Nature of Modified Teleparallel Theories}},}\
  }\href@noop {} {\  (\bibinfo {year} {2017})},\ \Eprint
  {http://arxiv.org/abs/1705.01072} {arXiv:1705.01072 [gr-qc]} \BibitemShut
  {NoStop}%
\bibitem [{\citenamefont {Hohmann}\ \emph {et~al.}(2017)\citenamefont
  {Hohmann}, \citenamefont {Järv}, \citenamefont {Krššák},\ and\
  \citenamefont {Pfeifer}}]{Hohmann:2017duq}%
  \BibitemOpen
  \bibfield  {author} {\bibinfo {author} {\bibfnamefont {Manuel}\ \bibnamefont
  {Hohmann}}, \bibinfo {author} {\bibfnamefont {Laur}\ \bibnamefont {Järv}},
  \bibinfo {author} {\bibfnamefont {Martin}\ \bibnamefont {Krššák}}, \ and\
  \bibinfo {author} {\bibfnamefont {Christian}\ \bibnamefont {Pfeifer}},\
  }\bibfield  {title} {\enquote {\bibinfo {title} {{Teleparallel theories of
  gravity as analogue of non-linear electrodynamics}},}\ }\href@noop {} {\
  (\bibinfo {year} {2017})},\ \Eprint {http://arxiv.org/abs/1711.09930}
  {arXiv:1711.09930 [gr-qc]} \BibitemShut {NoStop}%
\bibitem [{\citenamefont {Geng}\ \emph {et~al.}(2011)\citenamefont {Geng},
  \citenamefont {Lee}, \citenamefont {Saridakis},\ and\ \citenamefont
  {Wu}}]{Geng:2011aj}%
  \BibitemOpen
  \bibfield  {author} {\bibinfo {author} {\bibfnamefont {Chao-Qiang}\
  \bibnamefont {Geng}}, \bibinfo {author} {\bibfnamefont {Chung-Chi}\
  \bibnamefont {Lee}}, \bibinfo {author} {\bibfnamefont {Emmanuel~N.}\
  \bibnamefont {Saridakis}}, \ and\ \bibinfo {author} {\bibfnamefont {Yi-Peng}\
  \bibnamefont {Wu}},\ }\bibfield  {title} {\enquote {\bibinfo {title}
  {{“Teleparallel” dark energy}},}\ }\href {\doibase
  10.1016/j.physletb.2011.09.082} {\bibfield  {journal} {\bibinfo  {journal}
  {Phys. Lett.}\ }\textbf {\bibinfo {volume} {B704}},\ \bibinfo {pages}
  {384--387} (\bibinfo {year} {2011})},\ \Eprint
  {http://arxiv.org/abs/1109.1092} {arXiv:1109.1092 [hep-th]} \BibitemShut
  {NoStop}%
\bibitem [{\citenamefont {Chakrabarti}\ \emph {et~al.}(2017)\citenamefont
  {Chakrabarti}, \citenamefont {Said},\ and\ \citenamefont
  {Farrugia}}]{Chakrabarti:2017moe}%
  \BibitemOpen
  \bibfield  {author} {\bibinfo {author} {\bibfnamefont {Soumya}\ \bibnamefont
  {Chakrabarti}}, \bibinfo {author} {\bibfnamefont {Jackson~Levi}\ \bibnamefont
  {Said}}, \ and\ \bibinfo {author} {\bibfnamefont {Gabriel}\ \bibnamefont
  {Farrugia}},\ }\bibfield  {title} {\enquote {\bibinfo {title} {{Some aspects
  of reconstruction using a scalar field in $f(T)$ gravity}},}\ }\href
  {\doibase 10.1140/epjc/s10052-017-5404-6} {\bibfield  {journal} {\bibinfo
  {journal} {Eur. Phys. J.}\ }\textbf {\bibinfo {volume} {C77}},\ \bibinfo
  {pages} {815} (\bibinfo {year} {2017})},\ \Eprint
  {http://arxiv.org/abs/1711.04423} {arXiv:1711.04423 [gr-qc]} \BibitemShut
  {NoStop}%
\bibitem [{\citenamefont {Otalora}(2013)}]{Otalora:2013tba}%
  \BibitemOpen
  \bibfield  {author} {\bibinfo {author} {\bibfnamefont {G.}~\bibnamefont
  {Otalora}},\ }\bibfield  {title} {\enquote {\bibinfo {title} {{Scaling
  attractors in interacting teleparallel dark energy}},}\ }\href {\doibase
  10.1088/1475-7516/2013/07/044} {\bibfield  {journal} {\bibinfo  {journal}
  {JCAP}\ }\textbf {\bibinfo {volume} {1307}},\ \bibinfo {pages} {044}
  (\bibinfo {year} {2013})},\ \Eprint {http://arxiv.org/abs/1305.0474}
  {arXiv:1305.0474 [gr-qc]} \BibitemShut {NoStop}%
\bibitem [{\citenamefont {Jamil}\ \emph {et~al.}(2012)\citenamefont {Jamil},
  \citenamefont {Momeni},\ and\ \citenamefont {Myrzakulov}}]{Jamil:2012vb}%
  \BibitemOpen
  \bibfield  {author} {\bibinfo {author} {\bibfnamefont {Mubasher}\
  \bibnamefont {Jamil}}, \bibinfo {author} {\bibfnamefont {D.}~\bibnamefont
  {Momeni}}, \ and\ \bibinfo {author} {\bibfnamefont {Ratbay}\ \bibnamefont
  {Myrzakulov}},\ }\bibfield  {title} {\enquote {\bibinfo {title} {{Stability
  of a non-minimally conformally coupled scalar field in F(T) cosmology}},}\
  }\href {\doibase 10.1140/epjc/s10052-012-2075-1} {\bibfield  {journal}
  {\bibinfo  {journal} {Eur. Phys. J.}\ }\textbf {\bibinfo {volume} {C72}},\
  \bibinfo {pages} {2075} (\bibinfo {year} {2012})},\ \Eprint
  {http://arxiv.org/abs/1208.0025} {arXiv:1208.0025 [gr-qc]} \BibitemShut
  {NoStop}%
\bibitem [{\citenamefont {Chen}\ \emph
  {et~al.}(2015{\natexlab{b}})\citenamefont {Chen}, \citenamefont {Wu},\ and\
  \citenamefont {Wei}}]{Chen:2014qsa}%
  \BibitemOpen
  \bibfield  {author} {\bibinfo {author} {\bibfnamefont {Zu-Cheng}\
  \bibnamefont {Chen}}, \bibinfo {author} {\bibfnamefont {You}\ \bibnamefont
  {Wu}}, \ and\ \bibinfo {author} {\bibfnamefont {Hao}\ \bibnamefont {Wei}},\
  }\bibfield  {title} {\enquote {\bibinfo {title} {{Post-Newtonian
  Approximation of Teleparallel Gravity Coupled with a Scalar Field}},}\ }\href
  {\doibase 10.1016/j.nuclphysb.2015.03.012} {\bibfield  {journal} {\bibinfo
  {journal} {Nucl. Phys.}\ }\textbf {\bibinfo {volume} {B894}},\ \bibinfo
  {pages} {422--438} (\bibinfo {year} {2015}{\natexlab{b}})},\ \Eprint
  {http://arxiv.org/abs/1410.7715} {arXiv:1410.7715 [gr-qc]} \BibitemShut
  {NoStop}%
\bibitem [{\citenamefont {Hohmann}\ \emph
  {et~al.}(2018{\natexlab{a}})\citenamefont {Hohmann}, \citenamefont {Järv},\
  and\ \citenamefont {Ualikhanova}}]{Hohmann:2018rwf}%
  \BibitemOpen
  \bibfield  {author} {\bibinfo {author} {\bibfnamefont {Manuel}\ \bibnamefont
  {Hohmann}}, \bibinfo {author} {\bibfnamefont {Laur}\ \bibnamefont {Järv}}, \
  and\ \bibinfo {author} {\bibfnamefont {Ulbossyn}\ \bibnamefont
  {Ualikhanova}},\ }\bibfield  {title} {\enquote {\bibinfo {title} {{Covariant
  formulation of scalar-torsion gravity}},}\ }\href@noop {} {\  (\bibinfo
  {year} {2018}{\natexlab{a}})},\ \Eprint {http://arxiv.org/abs/1801.05786}
  {arXiv:1801.05786 [gr-qc]} \BibitemShut {NoStop}%
\bibitem [{\citenamefont {Hohmann}\ and\ \citenamefont
  {Pfeifer}(2018)}]{Hohmann:2018dqh}%
  \BibitemOpen
  \bibfield  {author} {\bibinfo {author} {\bibfnamefont {Manuel}\ \bibnamefont
  {Hohmann}}\ and\ \bibinfo {author} {\bibfnamefont {Christian}\ \bibnamefont
  {Pfeifer}},\ }\bibfield  {title} {\enquote {\bibinfo {title} {{Scalar-torsion
  theories of gravity II: $L(T, X, Y, \phi)$ theory}},}\ }\href@noop {} {\
  (\bibinfo {year} {2018})},\ \Eprint {http://arxiv.org/abs/1801.06536}
  {arXiv:1801.06536 [gr-qc]} \BibitemShut {NoStop}%
\bibitem [{\citenamefont {Hohmann}(2018)}]{Hohmann:2018ijr}%
  \BibitemOpen
  \bibfield  {author} {\bibinfo {author} {\bibfnamefont {Manuel}\ \bibnamefont
  {Hohmann}},\ }\bibfield  {title} {\enquote {\bibinfo {title} {{Scalar-torsion
  theories of gravity III: analogue of scalar-tensor gravity and conformal
  invariants}},}\ }\href@noop {} {\  (\bibinfo {year} {2018})},\ \Eprint
  {http://arxiv.org/abs/1801.06531} {arXiv:1801.06531 [gr-qc]} \BibitemShut
  {NoStop}%
\bibitem [{\citenamefont {Hohmann}(2016)}]{Hohmann:2015pva}%
  \BibitemOpen
  \bibfield  {author} {\bibinfo {author} {\bibfnamefont {Manuel}\ \bibnamefont
  {Hohmann}},\ }\bibfield  {title} {\enquote {\bibinfo {title} {{Spacetime and
  observer space symmetries in the language of Cartan geometry}},}\ }\href
  {\doibase 10.1063/1.4961152} {\bibfield  {journal} {\bibinfo  {journal} {J.
  Math. Phys.}\ }\textbf {\bibinfo {volume} {57}},\ \bibinfo {pages} {082502}
  (\bibinfo {year} {2016})},\ \Eprint {http://arxiv.org/abs/1505.07809}
  {arXiv:1505.07809 [math-ph]} \BibitemShut {NoStop}%
\bibitem [{\citenamefont {Hohmann}\ \emph
  {et~al.}(2018{\natexlab{b}})\citenamefont {Hohmann}, \citenamefont {Järv},
  \citenamefont {Krššák},\ and\ \citenamefont {Pfeifer}}]{Hohmann:2018sym}%
  \BibitemOpen
  \bibfield  {author} {\bibinfo {author} {\bibfnamefont {Manuel}\ \bibnamefont
  {Hohmann}}, \bibinfo {author} {\bibfnamefont {Laur}\ \bibnamefont {Järv}},
  \bibinfo {author} {\bibfnamefont {Martin}\ \bibnamefont {Krššák}}, \ and\
  \bibinfo {author} {\bibfnamefont {Christian}\ \bibnamefont {Pfeifer}},\
  }\bibfield  {title} {\enquote {\bibinfo {title} {{Modified teleparallel
  theories of gravity in symmetric spacetimes}},}\ }\href@noop {} {\  (\bibinfo
  {year} {2018}{\natexlab{b}})},\ \bibinfo {note} {to appear}\BibitemShut
  {NoStop}%
\bibitem [{\citenamefont {Baez}\ and\ \citenamefont
  {Wise}(2015)}]{Baez:2012bn}%
  \BibitemOpen
  \bibfield  {author} {\bibinfo {author} {\bibfnamefont {John~C.}\ \bibnamefont
  {Baez}}\ and\ \bibinfo {author} {\bibfnamefont {Derek~K.}\ \bibnamefont
  {Wise}},\ }\bibfield  {title} {\enquote {\bibinfo {title} {{Teleparallel
  Gravity as a Higher Gauge Theory}},}\ }\href {\doibase
  10.1007/s00220-014-2178-7} {\bibfield  {journal} {\bibinfo  {journal}
  {Commun. Math. Phys.}\ }\textbf {\bibinfo {volume} {333}},\ \bibinfo {pages}
  {153--186} (\bibinfo {year} {2015})},\ \Eprint
  {http://arxiv.org/abs/1204.4339} {arXiv:1204.4339 [gr-qc]} \BibitemShut
  {NoStop}%
\end{thebibliography}%
\end{document}